\documentstyle{aipproc2}
\input psfig.sty
\input epsf
\epsfverbosetrue
\def\etal   {{\rm et~al.}}

\def\wisk#1{\ifmmode{#1}\else{$#1$}\fi}

\def\yff{{Y_{\rm ff}}}
\def\percm  {\wisk{{\rm cm}^{-1}}}

\def\lsim   {\wisk{_<\atop^{\sim}}}
\def\gsim   {\wisk{_>\atop^{\sim}}}

\def\deg    {\wisk{^\circ}}

\def\muK{\wisk{{\rm \mu K}}}

\def\apj{Astrophysical Journal}

\begin{document}

\title{COBE Observations and Results}

\author{George F. Smoot}
\address{Lawrence Berkeley National Laboratory\\
Space Sciences Laboratory\\
Center for Particle Astrophysics\\
Department of Physics\\
University of California, Berkeley CA 94720\\
e-mail: Smoot@cosmos.lbl.gov }

\maketitle

\begin{abstract}
This paper summarizes the results from the COBE satellite mission.
Nine years have passed since the launch of COBE and six years 
since the announcement of the discovery of cosmic 
microwave background anisotropies by the COBE DMR instrument.
This is still a relatively short time to look back and 
understand the implications of COBE and the anisotropy discovery; 
however, this 3K Cosmology Conference provides some context.
The Cosmic Background Explorer (COBE) satellite has made a major contribution
to the field of cosmology and has help create the confidence and high
level of interest that propels the field today.
Two major CMB observations, the thermal spectrum of the CMB and the CMB
anisotropies, plus a host of other observations and conclusions
are the basis and a major but not exclusive portion of the legacy of COBE.
The recent detection and observation of the cosmic infrared background (CIB)
are also part of COBE's major contribution to cosmology.
\end{abstract}

\noindent
{\it Subject headings:} cosmic microwave background -- cosmology --
-- artificial satellites, space probes

\section*{Introduction}
COBE (the Cosmic Background Explorer satellite) is NASA's first cosmological
satellite. 
The spectacular results and success COBE provided stimulation to the field and
to everyone's imagination which has resulted in the world's space agencies 
being open to future missions that promise to revolutionize and 
quantify our view of the Universe.
COBE has also provided us with confidence in the Big Bang model of cosmology
and our ability to test variants of the Big Bang models and determine
their parameters with precision.
COBE also determined two of those parameters remarkable well in a field
characterized by uncertainty in these parameters at the 50 to 100\%\ level.

The Cosmic Background Explorer (COBE)
was NASA's first cosmological satellite
and had a number of significant results:
\begin{itemize}

\item Full Sky Maps at 14 Frequencies

\item Cosmic Microwave Background Spectrum -- precision measurements

\item CMB Anisotropies -- detection, observation, measurement

\item Diffuse Infrared Background -- detection, measurement

\item CMB Polarization -- significant upper limits

\end{itemize}

COBE, additionally, proved to be a training ground for approaches, techniques,
and people. These included those of the COBE team and the external community
which had good and timely access to the data. This lead to much scientific
interest and activity. 
Our current view of the data/signal flow and ultimate analysis and 
interpretation
is very similar to that used on the COBE observations and analysis.

Boggess et al.\ \cite{Boggess92} provide a COBE mission overview.

\section{CMB Thermal Spectrum}

\subsection{Frequency Spectrum}

The FIRAS (Far Infrared Absolute Spectrophotometer)
instrument has made a high precision observation
of the cosmic microwave background.

FIRAS \cite{Mather90}  was built to measure
the spectrum of diffuse emission from 1 to 100 \percm, with particular
attention to possible differences between the spectrum of the
cosmic microwave background radiation (CMBR) and a blackbody
spectrum as small as 0.1\%\ of the peak of the CMBR spectrum.
The FIRAS  has differential inputs and outputs, 
a full beam external calibrator, a controllable reference blackbody, 
and a polarizing Michelson interferometer with bolometer detectors.
It operated at a temperature of 1.5 K inside a liquid helium cryostat
to suppress instrument emission and improve detector sensitivities.

FIRAS improves upon previous measurements in reducing the
potential contributions of the systematic errors by: 
1) operating outside the atmosphere;  
2) providing full aperture {\it in situ} calibration; 
3) providing a continuous differential comparison with a reference blackbody
adjusted to null the input signal; 
4) operating the entire instrument, including the beam forming optics, 
in a shielded environment at cryogenic temperatures; and 
5) using an improved horn antenna with a flared aperture to
define the beam and reduce the contributions from objects outside
the main beam.
These were possible in the context of a satellite mission.

\begin{figure}[tb]
\psfig{figure=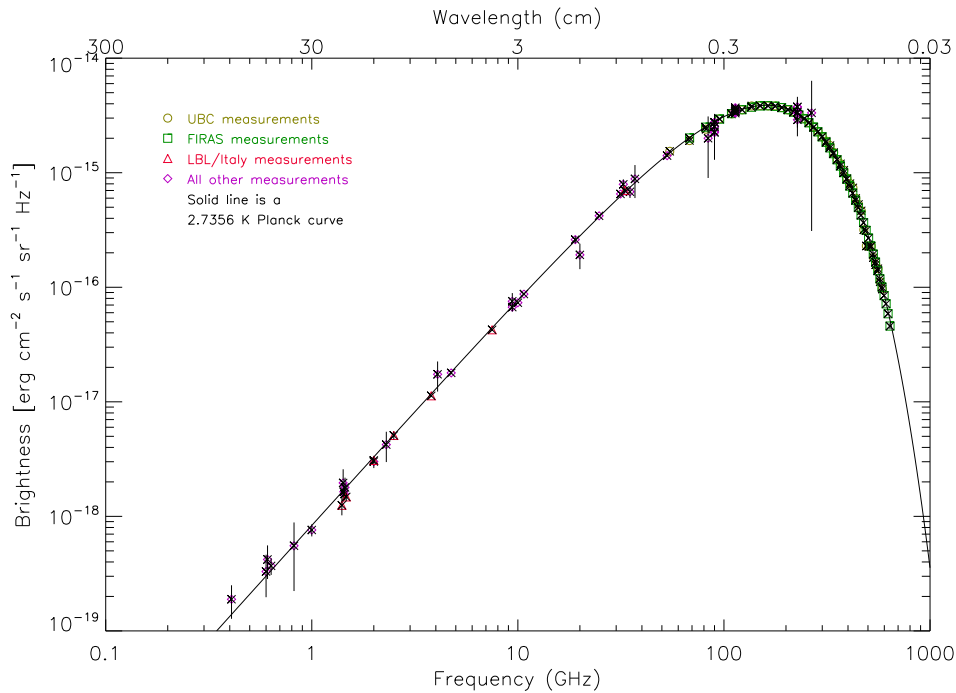,angle=90.,height=5.0in}
\caption{Brightness as a function of frequency of the CMB
\label{fig:spectrum}}
\end{figure}

\begin{figure}[tb]
\psfig{figure=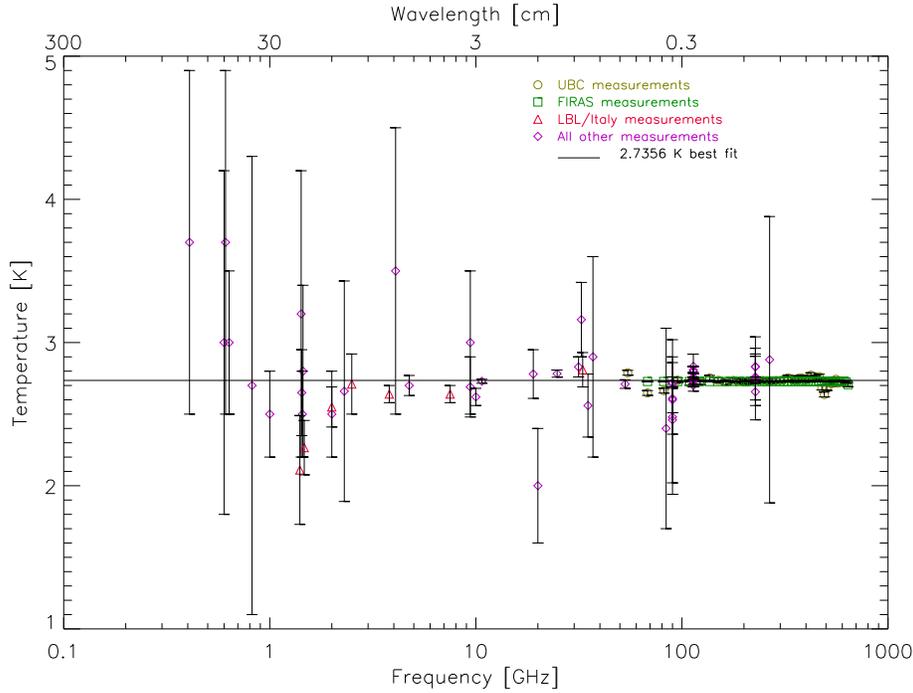,angle=90.,height=5.0in}
\caption{Thermodynamic temperature as a function of frequency of the CMB
\label{fig:tspectrum}}
\end{figure}

The ultimate result of the FIRAS observations and analysis 
was a remarkable precise measurement of the CMB spectrum
and comparison \cite{fixsen96} to that from a blackbody spectrum.
The measured spectrum is shown in Figure \ref{fig:spectrum}
and \ref{fig:tspectrum}.

The CMB spectrum would have a blackbody form if the simple, hot, Big Bang
model is a correct description of the early universe,
but will be distorted from that form by energy release
for a redshift $z \lsim3\times10^6$ \cite{sunyaev80}. 
Such releases might arise from decay of unstable particles, dissipation of
cosmic turbulence and gravitational waves, breakdown of cosmic strings and
other more exotic transformations.
The CMB was the dominant energy field after the annihilation of positrons
and the decoupling of neutrinos until
$ z \gsim 3\times 10^4$. 
For energy release into the electron-proton plasma, between redshift of
a few times $10^4$ and $10^6$ \cite{smoot88}
the number of Compton scatterings is sufficient to bring the photons
into thermal equilibrium with the primordial plasma.
During this time, bremsstrahlung and other radiative processes do not have
enough time to add a sufficient amount of photons to create a
Planckian distribution.
The resulting distribution is a Bose-Einstein spectrum
with a chemical potential, $\mu$, that is exponentially attenuated
at low frequencies, $\mu = \mu _0 e^{-2\nu_b /\nu}$
(where $\nu_b$ is the frequency at which Compton scattering of photons
to higher frequencies is balanced by bremsstrahlung creation
at lower frequencies \cite{danese82}).
At redshifts smaller than $\sim10^5$,
the Compton scattering rate is no longer high
enough to produce a Bose-Einstein spectrum.
The resulting spectrum has an increased brightness temperature in the far
Rayleigh-Jeans region due to bremsstrahlung emission by relatively hot
electrons,
a reduced temperature in the middle Rayleigh-Jeans region where the photons are
depleted by Compton scattering, and a high temperature in the Wien region where
the Compton-scattered photons have accumulated.

These observations provide new limits on the following cosmological
parameters: \newline
$T_0$ -- the thermodynamic temperature of the background radiation     \newline
$\yff$ -- a measure of the effects of Bremsstrahlung, unitless; a.k.a.
${J_{\rm ff}}$ \newline
y -- a measure of the effects of Compton scattering, unitless   \newline
$\mu_0$ -- describes the Bose-Einstein distortion, unitless.    \newline

The most complete fit is a nonlinear simultaneous fit of $T_0$,
$\yff$, $\mu_0$, $y$ and $\beta$, as shown in Table \ref{tab:nonlinear},
making use of all data published for the CMB monopole. 
\begin{table*}
\caption{Non-linear fit with $T_0$, $\yff$, $\mu_0$ and $y$
to CMB spectrum monopole.} 
\begin{center}
\begin{tabular}{llrcll}
$T_0     $&=&$  2.7377 $ & $\pm$ & $  0.0038 $ K &(95\% CL)\\
$\yff    $&=&$ -1.1\times10^{-5}     $&$\pm$&$  2.3\times10^{-5} $ &(95\% CL)\\
$\mu_0   $&=&$ -3.0\times10^{-5}     $&$\pm$&$  1.2\times10^{-4} $ &(95\% CL)\\
$y       $&=&$  1.6\times10^{-6}     $&$\pm$&$  9.6\times10^{-6} $ &(95\% CL)\\
\end{tabular}
\label{tab:nonlinear}
\end{center}
\begin{center}
 Correlation matrix = \hspace{-.15in} 
\begin{math}
        \begin{array}{c} \\ T_0 \\ \yff \\ \mu_0 \\ y \end{array} \hspace{-.1in}
        \begin{array}{c} \\ \left[ \hspace{-.25in} \begin{array}{c} \\ \\ \\ \\
\end{array} \right. \end{array}
        \begin{array}{rrrr}
  T_0& \yff& \mu_0 & y \\
   1.00 & 0.01 & 0.06 & 0.09 \\
   0.01 & 1.00 &-0.27 &-0.18 \\
   0.06 &-0.27 & 1.00 & 0.82 \\
   0.09 &-0.18 & 0.82 & 1.00 \\
        \end{array}
        \begin{array}{c} \\ \left. \begin{array}{c} \\ \\ \\ \\ \end{array}
\hspace{-.15in} \right] \end{array}
\end{math}
\end{center}
\end{table*}

Big Bang only theory to give this precisely thermal (Planckian) spectrum.

\subsection{Energy Limits}

Energy released at different epochs
probes different physical conditions in the early universe
and creates different signatures in the CMB spectrum.
Figure \ref{fig:spectrum} shows the current spectrum observations
and sample spectral distortion
characteristic of each mechanism.
Energy release at recent epochs ($z < 1000$)
will re-ionize the intergalactic medium,
which then cools through free-free emission.
If the gas is hot enough or the release occurs before recombination
($z < 10^5$), Compton scattering of CMB photons
from hot electrons provides the primary cooling mechanism.
Early energy release ($10^5 < z < 10^7$) from relic decay
reaches statistical equilibrium,
characterized by a chemical potential distortion at long wavelengths.

In general the observational limits on potential spectral distortions
in turn limit the energy release in the universe to roughly $10^{-4}$
of the energy of the cosmic background radiation for redshifts
between $10^3 < z < 10^7$.

The greatest significance of the non-detection of distortions
in the CMB spectrum is the support it gives to
the Hot Big Bang model of cosmology.
A blackbody spectrum is the natural result of a system in thermal equilibrium;
conversely, it is very difficult to superpose a set of non-thermal spectra
or thermal spectra of different temperatures
to mimic a blackbody to such tight tolerance over so broad a frequency range.
Steady-state models generally create a microwave background
by absorption and subsequent re-emission of starlight by dust.
The re-emitted photons are then redshifted by the expansion of the universe,
so that we observe a superposition of ``shells'' radiating at different epochs.
If the dust opacity were independent of frequency,
the microwave background would be a superposition
of thermal spectra redshifted to different temperatures,
equivalent to a Compton $y$ distortion.
The dust opacity can be tailored to bring the superposition
closer to a blackbody,
but the required (very high) opacities at millimeter wavelengths
conflict with direct observations
of high-redshift galaxies at these wavelengths (\cite{peebles93} p203-6).

The same basic arguments rule out ``plasma universes'' models.

The lack of spectral distortions has implications for structure formation.
Explosive models of structure formation
provide a simple way to sweep (baryonic) matter out of the observed voids.
The resulting shocks, however, heat the baryons and distort the CMB spectrum.
Upper limits on a Compton $y$ distortion
limit the maximum size of explosive voids to
$R_{\rm void} < 2$ Mpc,
much smaller than the observed voids \cite{levin92}.

\subsection{Dipole Anisotropy Spectrum}

COBE FIRAS (\& DMR) measured the spectrum of the CMB monopole and dipole
with unprecedented accuracy.
These observations have shown that the dipole frequency spectrum
matches the differential to a Planckian to high precision.
This is a cornerstone in our understanding of the expected
frequency spectrum of the CMB anisotropies, which is less well
measured by COBE, and thus the ability to relate measurements
at different frequencies and to separate out potential foregrounds
by their differing frequency spectra.

This provides:
\begin{itemize}

\item Support for Big Bang and Thermal Origin of the CMB

\item $T_{CMB}$, $n_\gamma$, $v_{peculiar}$

\item Spectral Shape for Anisotropies versus foregrounds

\item Calibration for other experiments

\item Limits on distortions and Energy Release in early universe

\end{itemize}

\subsection{Dipole as Check and Calibration (1\%)}
As observers we have been fortunate to have the dipole anisotropy
as a reasonable signal on which to check and calibrate
our anisotropy observations.
COBE provides us with a sufficiently good observation and accurate
measurement of the dipole both from its own calibration
and the having observed the dipole and its annual modulation
through four seasons.
This provides a sufficiently accurate direction and amplitude\cite{lineweaver96}
of the dipole to make it a reference standard for succeeding experiments.

\section{CMB Anisotropies}

COBE DMR detected and measured CMB anisotropies 
\begin{itemize}

\item First detection of primordial CMB anisotropies

\item Determination of Primordial Fluctuation Normalization at $10^{-5}$

\item Still best determination, cosmic variance limited, 
of large angular scale power spectrum

\item With FIRAS good determination anisotropies have proper frequency spectrum

\item Information on the Large Scale Structure of Space-Time

\item Scientifically useful test-bed data set and procedures

\end{itemize}

A very major result from COBE was the DMR detection and reporting
of CMB anisotropies.
CMB anisotropies had long been sought and it is easy to forget
how with year after year of improving upper limits
on anisotropies people in the field had got to the mind set
that there were no anisotropies and that quality
of an experiment was in how low a limit it could set.
A prime reason for the invention of cold dark matter
was to avoid potential violation of CMB anisotropy limits.
Things had gotten to the state that critics of the Big Bang cosmology
were pointing to the lack of CMB anisotropies as a indicator 
of Big Bang model failure.

With the DMR announcement of the discovery of CMB anisotropies
on all observed angular scales and with a near-scale invariant spectrum,
the field was invigorated both by the results and by the great public interest.
What soon became known as the COBE normalization was just
a factor of two above the minimum required for gravitational instability
formation of large scale structure.
That plus the angular power spectrum was just what the theorists
had ordered and gave the push to what turned out to be 
a rapid theoretical development of CMB anisotropy theory
and related fields.

The response on the experimental/observational side was also
quite impressive with new observations and instrumental developments
occurring in rapid succession.
Governmental agencies (e.g. NASA and NSF) gave high priority
and significant funding support to CMB science.
The most impressive long-term exploitation of the CMB were the development
of proposed CMB anisotropy satellite by teams of scientists.
These lead to NASA and ESA accepting the MAP and Planck missions.

The CMB power spectrum can readily be interpreted for what it tells us
about the primordial power spectrum of fluctuations
and then applied either backwards
to tell us about early-universe, high-energy physics or viewed forward
in its relation to large scale structure.
The field, especially theoretically and rightly so, is focused on these issues.
However, there are other things that we can learn about the universe
using these data. We can learn a lot about the geometry and dynamics
of the universe and set very tight limits on anisotropic Hubble expansion
and on shear and rotation (vorticity) in the universe.
We also know that the geometry of the universe is very near to the
idealized Robertson-Walker metric with only small perturbations
which must be on the same scale as the CMB anisotropies.
In 1968 Ehlers, Geren, and Sachs proved the theorem: If a family of
freely-falling observers measure self-gravitating background radiation
to be {\it exactly} isotropic, then the Universe is {\it exactly}
Friedmann-Lemaitre-Robertson-Walker. This has been interpreted/extrapolated
by me and others that the observed high degree of observed CMB isotropy
and the assumption that our location is not special imply that
the Universe's metric is nearly Robertson-Walker with perturbations
of the order of $10^{-5}$.
After COBE Stoeger, Maartens, and Ellis have shown that if all fundamental
observers measure the cosmic microwave background radiation to be
almost isotropic in an expanding Universe, or that one observer
observing over all time sees a nearly isotropic CMB, then that Universe
is almost spatially homogeneous and isotropic. This puts us on
firmer footing and allows us to use the CMB isotropy observations
to set limits on the anisotropy, homogeneity, and dynamics of the Universe.

The fact that the CMB anisotropies show that we can treat the Universe
as if it has a Robertson-Walker metric with small perturbations,
justifies the assumption that large scale structure theory had been 
built upon as it simplifies and makes the mathematics tractable.

The DMR consists of 6 differential microwave radiometers: 2 nearly independent
channels, labeled A and B, at frequencies 31.5, 53, and 90 GHz (wavelength
9.5, 5.7, and 3.3 mm respectively).
Each radiometer measures the difference in power between two 7$^\circ$ fields
of view separated by 60$^\circ$, 30$^\circ$ to either side
of the spacecraft spin axis \cite{Smoot90}.
{\it COBE} was launched from Vandenberg Air Force Base on 18 November 1989
into a 900 km, 99$^\circ$ inclination circular orbit,
which precesses to follow the terminator (light dark line on the Earth)
as the Earth orbits the Sun.
Attitude control keeps the spacecraft pointed away from the Earth and nearly
perpendicular to the Sun with a slight backward tilt so that solar radiation
never directly illuminates the aperture plane.
The combined motions of the spacecraft spin (75 s period),
orbit (103 m period), and orbital precession ($\sim 1^\circ$ per day) allow
each sky position to be compared to all others through a highly redundant set
of temperature difference measurements spaced 60$^\circ$ apart.
The on-board processor box-car integrates the differential signal
from each channel for 0.5 s, and records the digitized differences
for daily playback to a ground station.

\subsection{COBE DMR 4-year Maps}
The COBE Differential Microwave Radiometers (DMR) instrument
has mapped the full microwave sky 
to sensitivity 17 \muK ~per 7\deg ~field of view.

The maps are of high quality and provide redundancy of observations.
For example the 53 GHz 4-year A+B map has signal-to-noise ratio is larger 
than two.

The observation of anisotropies opens one new cosmological test: probing
topology of the Universe. The most straightforward topology that we can imagine
is an essentially isotropic and homogeneous Universe that is simply connected
However, we know of no constraints that actually require that the Universe
be simply connected. It might in fact have the topology of a donut
or many other possible objects.

Over time there have been reports of periodicity in the Universe
both in terms of large scale structure of galaxies ($\sim$128 Mpc)
and quasars (much larger scale). This has led a number of people
to suggest that the Universe is small with opposite faces identified
(or some other combination). Such universes would be periodic on
the identified axes and thus could not have anisotropies with wavelengths
longer than their symmetry axes.
The existence of very large angular scale anisotropies, i.e. the quadrupole,
octupole, and hexadecapole, put stringent limits on the size of the universe.
Various analyses of the DMR data set a limit on the smallness of the universe
at about 0.5 of the Hubble diameter (e.g. \cite{Jing94}, \cite{Costa94}).
This is an illustration of the power of the CMB as a cosmological probe.

\subsection{Ability to Make Observations with Sufficiently Low Systematics}
A key issue in making observations at such low signal levels
is whether systematic errors are likely to provide fundamental limits.
A a detailed treatment of the upper
limits on residual systematic errors (\cite{Kogut92}, \cite{Kogut96a})
by the DMR team showed convincingly that one could
reduce the systematic errors to well below the 10\%\ level
in the temperature maps and the 1\%\ level in the power spectrum.
This was further enhanced by the cross-correlation of the DMR maps
with other results as additional new observations were made.
In particular the positive
cross-correlation between the {\it COBE} data and data from balloon-borne
observations at a shorter wavelength \cite{Ganga93}
and later by comparison of the {\it COBE} data and data from the ground-based
Tenerife experiment \cite{Lineweaver95} at longer wavelengths
showed that the anisotropy features were real and had the expected
spectral dependence.
This conclusion was based upon the long lever arm and the realization that the 
COBE DMR observation frequencies were essentially centered 
in a low valley between Galactic and extra-galactic foreground signals.

COBE DMR provided a crucial test of how well an experiment could control
the systematics as the circumstances, frequency and hardware redundancy,
and careful analysis of the observations were provided under the aegis
of a satellite mission.

\subsection{Treatment and Level of Foregrounds}
The correlation between ``free-free'' and dust emissions was 
demonstrated with the DMR data. 
This was a rediscovery of the known partial correlation between
HII regions and dust 
but also lead to the realization that there might be other 
forms of emission such as that from rotating dust grains.

As COBE was able to separate foregrounds and CMB signals reasonably well,
and thus was also able to measure the power spectrum of anisotropies versus 
foregrounds (e.g. \cite{Gorski96}).
This lead to the possibility that the CMB and foregrounds might
also be distinguished by their power spectra as well as frequency spectra.

This also lead to a set of programs to understand the possible confusing 
foregrounds and find methods to separate them. 
Key examples include the Galactic Emission Mapping (GEM) project
(URL http://aether.lbl.gov/www/projects/gem) discussed in this talk,
the WOMBAT challenge (URL http://astro.berkeley.edu/wombat/), 
and the MAP and Planck efforts.

\subsection{Gaussian versus non-Gaussian}
The COBE DMR was the first experiment to offer up data and analysis
to show that the CMB anisotropies are consistent with Gaussian
fluctuations and show that there was little evidence of non-Gaussianity
(Only two papers by any one analyzing the data claim any evidence
of non-gaussianity and that evidence is very restricted.)
Gaussianity is a standard assumption of many analysis techniques
and most models.

\subsection{Summary of 4-Year COBE-DMR CMB Measurements}

\begin{itemize}

\item Consistent with results from first- and two-year data

\item Signal-to-noise ratio in a 10$^\circ$ smoothed map is 2,
enough to provide a clear visual impression of the data.

\item CMB monopole temperature is $T_0 = 2.725 \pm 0.020$~K

\item CMB dipole amplitude is $3.353 \pm 0.024$~mK \\
~~~~toward 11 h $12.2 \pm 0.8$~m, -7.06$^\circ \pm 0.16^\circ$

\item CMB quadruple rms amplitude is $ 4 < Q_{rms} < 28 \mu K$
at the 95\%\ confidence level

\item CMB quadrupole amplitude fitted to a
power-law spectrum is $Q_{rms-PS} = 15.3^{+3.8}_{-2.8} ~\mu K$ \\
~~~~and to a sCDM model $17.5 \pm 2$.

\item The best-fitted power-law spectral index is $1.2 \pm 0.3$
Which is very close to what is expected from a scale-invariant
power spectrum of primordial fluctuations.

\item The DMR anisotropy data are consistent with Gaussian statistics.\\
Statistical tests prefer Gaussian to other models tested.


\end{itemize}

\subsection{4-Year COBE-DMR CMB Polarization Measurements}

Using the COBE DMR data we have been able to produce high quality
linear (53 \& 90 GHz) and circular polarization (31.5 GHz) maps of the 
full sky.

These maps show
\begin{itemize}

\item The degree of polarization is small. 
${\Delta P \over T} < {\Delta T \over T}$

\item Fits to global patterns show only upper limits:
    \begin{itemize}
	\item RMS polarization less than 10$^{-5}$	
 \item Temperature-polarization cross-correlation $<<$ temperature correlations
	\item Quadrupole-type polarization less than 10$^{-5}$.
	\item Combined polarization and temperature fit shows strong limit
on anisotropic expansion.
     \end{itemize}
\item Polarized signal from Galactic plane is small
\end{itemize}

The data are all consistent with no polarization at the 30\%
of the temperature anisotropy levels measured by the DMR
and thus with all standard models.
The 95\%\ C.L. limits on linear polarization amplitude is $P/T_{CMB} < 10^{-5}$
and on circular polarization are $V / T_{CMB} < 6 \times 10^{-5}$
on 7$^\circ$ and larger angular scales.
A combined fit to polarization and anisotropy limits anisotropic
axi-symmetric expansion to $\Delta H / H < 2 \times 10^{-6} $.

These are significant results on the polarization in and by themselves;
however, it is also instructive to compare them to theoretical predictions.
Nearly all theoretical calculations (e.g. \cite{Vittorio96}, \cite{Kam98})
indicate that the expected level of the polarized component is
of order 10\%\ of the temperature anisotopy for small angular scales
($> 1^\circ$) and significantly less for larger angular scales.
In general reionization results in a relatively larger ratio
but in no case would the polarized signal be larger than
the level of anisotropy and in most cases remains at the 10\%
or lower level.
Thus the DMR is consistent with the current theoretical predictions
but does not have the sensitivity to confirm them in a convincing manner.

\section{Diffuse Infrared Background}

A primary objective of the COBE mission has been the
detection and measurement of the diffuse cosmic infrared background (CIB).
The DIRBE (Diffuse Infrared Background Experiment) was designed
primarily to conduct a systematic search for an isotropic
infrared background in ten photometric bands
from 1.25 to 240 $\mu$m.

The search for the cosmic infrared background was an unfulfilled field
of observational cosmology.
The search for the CIB is impeded by two fundamental challenges:
no unique spectral signature was predicted for such a background
and there are many contributors locally and Galactically to the 
infrared sky brightness at all wavelengths.
Some of these sources, such as the interplanetary dust,
are quite bright.
The lack of unique spectral signature
arises in part because of the very many different
sources of such signals.
Possible existence of infrared backgrounds had been reported
by previous experiments but the community as a whole
remained far from convinced by the results.

The apparent attenuation of TeV gamma rays provided
indirect evidence for an infrared background \cite{Stecker97}.

Both the COBE DIRBE and FIRAS Instruments provide information
about the diffuse infrared background,
which at this point is primarily detected at long wavelengths.
Because of this the earliest reports came from the FIRAS data 
(see e.g. Puget et al. 1996).

The relevant papers by the COBE team are shown in Table \ref{tab:cib}.
\begin{table}
\caption{List of COBE team papers on CIB}
\label{tab:cib}
\vspace{0.4cm}
\begin{center}
\begin{tabular}{l l l l}
Author & Topic & Reference & astro-ph number \\
\hline
Hauser et al. (1998) & DIRBE Results Summary & Ap. J. 508, 25-43 &astro-ph/9806167 \\
Kelsall et al. (1998) & Zodiacal Light & Ap. J. 508, 44-73 & astro-ph/9806250 \\
Arendt et al. (1998) & Galactic Contribution & Ap. J. 508, 74-105 & astro-ph/9805323 \\
Dwek et al. (1998) & Cosmological Implications & Ap. J. 508, 106-122 & astro-ph/9806129 \\
Fixsen et al. (1998) & FIRAS: Spectrum of CIB & Ap. J. 508, 123-128 & astro-ph/9803021 \\
Dwek \& Arendt (1998) & 3.5 $\mu$m CIB & Ap. J. 508, L9-L12 & astro-ph/9809239 \\
\end{tabular} 
\end{center} 
\end{table}

The extragalactic background light consists of the cumulative emissions
from various pregalactic objects, protogalaxies, galaxies, and clusters
of galaxies summed over the evolution of the Universe.
Much of the extragalactic background light is predicted to be manifested
as the cosmic infrared background due to the process of absorption
and reradiation by dust particles and by the general redshifting
due to the expansion of the Universe.

Theorists have anticipated that there are two major energy sources
that contribute to the extragalactic background light: nuclear
and gravitational.
Nuclear energy released in stellar nucleosynthesis is radiated
predominantly in the ultraviolet to visible wavelengths and is
either redshifted or absorbed and reradiated into the infrared 
($\lambda \simeq 1 \mu$m) wavelength region.
Similarly, released gravitational energy is shifted toward longer wavelengths.
The infrared is thus expected to contain a significant fraction
of all the energy released in the Universe
since the recombination epoch.
Its observation then constrains the relative contribution
of all potential energy sources.

The integrated CIB intensity detected by COBE in the 140-1000~$\mu$m wavelength
range is about 16~nW~m$^{-2}$~sr$^{-1}$.
This intensity is consisentwith the energy release expected
from nuclear energy sources and xonstitutes about
20\% - 50\%\ of the total energy released in the formation of He
and metals throughout the history of the Universe.
Galaxy number counts provide a lower limit of 12~nW~m$^{-2}$~sr$^{-1}$ 
in the 0.36 to 2.2.~$\mu$m wavelength interval.
The explored regions then account for a total intensity of 
28~nW~m$^{-2}$~sr$^{-1}$.
If attributed to nuclear sources only,
this intensity implies more than about 10\%\ of the baryonic mass density
implied by big bang nucleosynthesis analysis
has been processed in stars to He and heavier elements.
This leaves little room for strong and exotic other sources of energy
release through this epoch.

\section{Summary/Perspective}

\begin{center}
{CONCLUSIONS: What We Learned from COBE?}
\end{center}

COBE demonstrated
\begin{itemize}

\item A Cosmology Satellite can be wildly Successful

\item A Development of Techniques and Personnel

\item CMB spectrum very close to Planckian

\item There are primordial perturbations / temperature anisotropies

\item Gravitational Instability is proper paradigm - normalization

\item The Hot Big Bang Model is strongly supported

\item The CMB polarization is small

\item The large-scale structure of space-time is simple

\item Constraints on non-standard physics to early times

\item CIB exists at an appropriate level

\end{itemize}

\acknowledgments
This work was funded in part by the Director, Office of Energy Research, 
Office of High Energy and Nuclear Physics, Division of High Energy Physics 
of the U.S. Department of Energy under contract DE-AC03-76SF00098.
We thank the 3 K cosmology conference for its support and efforts
under the leadership of Francesco Melchorri.

\end{document}